# A Scintillator Purification Plant and Fluid Handling System for SNO+

Richard J. Ford[1,*], for the SNO+ Collaboration

[1]*SNOLAB, Creighton Mine #9, 1039 R.R.24, Lively, Ontario, Canada.*

**Abstract.** A large capacity purification plant and fluid handling system has been constructed for the SNO+ neutrino and double-beta decay experiment, located 6800 feet underground at SNOLAB, Canada. SNO+ is a refurbishment of the SNO detector to fill the acrylic vessel with liquid scintillator based on Linear Alkylbenzene (LAB) and 2 g/L PPO, and also has a phase to load natural tellurium into the scintillator for a double-beta decay experiment with $^{130}$Te. The plant includes processes multi-stage dual-stream distillation, column water extraction, steam stripping, and functionalized silica gel adsorption columns. The plant also includes systems for preparing the scintillator with PPO and metal-loading the scintillator for double-beta decay exposure. We review the basis of design, the purification principles, specifications for the plant, and the construction and installations. The construction and commissioning status is updated.

## INTRODUCTION

The SNO+ experiment [1] is a re-task of the original Sudbury Neutrino Observatory detector [2] whereby the heavy water target is replaced with 780 tonnes of liquid scintillator based on Linear Alkylbenzene (LAB) and 2 g/L PPO wavelength shifter. The detector is located in SNOLAB [3] at an underground depth of 2,070 m at Vale Creighton Mine, near Sudbury, Ontario. The first phase of the experiment is to load the scintillator with telluric acid for double beta decay measurement with $^{130}$Te. Subsequently, SNO+ will have a rich experimental program for solar neutrino detection (in particular *pep* and CNO), geo-antineutrinos, reactor antineutrinos, supernovae and other exotic physics. The major detector construction was the installation of a hold-down rope net over the acrylic vessel (AV) and anchored to the cavity floor to counter the buoyancy of LAB, which has density 865 kg/m$^3$.

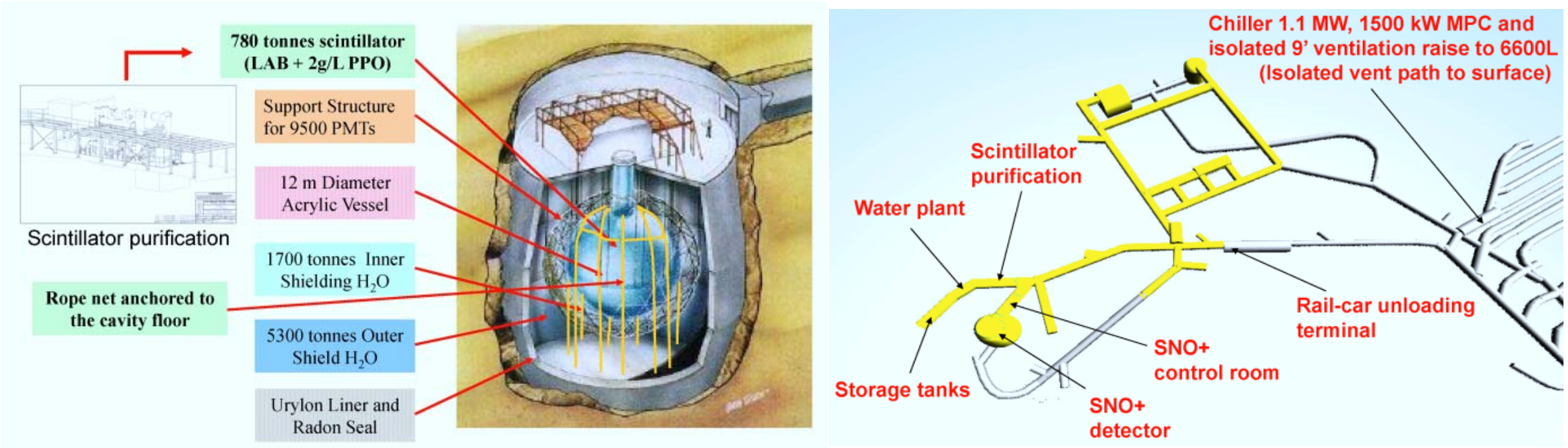

**FIGURE 1.** Sketch of the SNO+ detector and the major components, with the locations within SNOLAB.

* Corresponding author: email ford@snolab.ca.

The other major modification was the design and construction of a scintillator purification and liquid handling system, which is the topic of this paper. In this paper I discuss the process and plant design for the scintillator purification and liquid handling plants. The background requirements and reduction for the double beta decay phase, and the purification of the telluric acid, is not discussed here, as this is presented in another paper in these proceedings [4]. The SNO+ scintillator purity requirements, purification R&D, and specifications for the scintillator purification plant have been presented previously [5] and so are not discussed in detail here.

## INTERNAL BACKGROUNDS

The main internal backgrounds are from the $^{232}$Th chain and $^{238}$U chain daughters. One channel of contamination is from detector exposure to mine dust (which has ppm levels of U and Th), while another channel is U and Th inclusion in detector materials (AV, equipment, or piping) which allow mobile daughters such as radium, lead, and polonium to leach into the detector. Another contamination entry is radon exposure from air, leading to backgrounds from $^{214}$Bi, $^{210}$Pb, $^{210}$Bi, and $^{210}$Po. External backgrounds and analysis techniques are discussed in [1,4].

In consideration of all these backgrounds, the essential conclusion is that the LAB must be purified by distillation as the detector is filled, for removal of all U, Th, and K activities and to improve optical clarity. This initial fill purification flow rate should match the LAB delivery logistics, and must also include gas stripping to remove Rn, Ar, Kr and $O_2$. Subsequently the plant must have the capability to recirculate and repurify the scintillator for reduction as needed of leached and washed off Ra, Po, Bi and Pb, and for Rn due to Rn emanation and/or leakage to air. In the next section we discuss the purification methods and implementation within the plant.

## PURIFICATION METHODS AND PLANT DESIGN

Process and equipment design for the purification plant was highly challenging due to the location within a mine underground, and hazard controls for heating a combustible fluid. The process design was impacted by constraints on electrical power and cooling capacity, and limited liquid nitrogen that can be transported underground. Equipment design was severely limited by height availability for the columns, and equipment and piping density within the limited space available within the underground utility room. The purification plant comprises the processes of multi-stage distillation, solvent-solvent water extraction, gas stripping, and functional metal scavengers. Process engineering, including design of the columns, vessels, heat exchangers, and piping was done by Koch Modular Process Systems (KMPS) [6]. KMPS also provided the design of the purification system for Borexino [7].

### Multistage Distillation

Distillation is probably the oldest and best known of the separation processes, and is also the most effective in most cases. The process we are concerned with is continuous multi-stage column distillation (like an oil refinery) where the conditions are constant in time, which is distinct from batch distillation (eg. as for whiskey) where the components evolve over time. The feed is vapourized in the boiler and flows up a tower to be subsequently condensed as the purified product. The separation occurs due to difference in volatilities (ie. boiling point) and the process is best represented as distinct equilibrium stages. Liquid level remains in the boiler (bottoms) to concentrate the lower volatility components, and a small bottoms flow is discarded which removes the contaminants and maintains the bottoms concentration. The feed is introduced at a mid-point in the column, and will first flow down the column if the feed is sub-cooled with respect to the tower temperature. A fraction of the product is re-introduced sub-cooled into the top of the column, which is called the reflux. Above the feed are refining stages where heat is transferred from condensing contaminants to re-vaporize the more volatile components in the reflux. Below the feed is the stripping section which concentrates contaminants in the bottoms, which serves to reduce the required bottoms discard fraction. The optimal feed location balances the stripping and refining stages to obtain best product quality for fixed specification of the desired bottoms flow. For more detail the reader is referred to standard texts (eg. [8]).

Since the relative volatilities for heavy metals are very large, distillation effectively removes all radioactive heavy metals, in particular Ra, Th, Po, Pb, Bi, K. Distillation is the best method for improving the optical transmission of the scintillator, which is critical for increasing light collection to achieve the required detector resolution. Deteriorated optical quality typically results from oxygen exposure, and the impurities are assumed to be partially oxidized organic molecules, which will have lower volatility compared to pure LAB. Distillation will be used for the initial purification for the scintillator as part of the LAB delivery and detector filling process.

The column and process was designed by the engineering company KMPS, and utilized a detailed tower simulation. The basis of design was the separation of PPO, as proxy for all the unknown oxidized organics. We used PPO, as the relative volatility is not large ($\alpha$ = ~16) to ensure a conservative basis, and since separation of the PPO is a real requirement for decommissioning, and to enable the process design for bottoms distillation of PPO for scintillator distillation. The distillation is performed under vacuum, primarily for safety improvement as the atmospheric boiling point range of LAB (278–314°C) is close to the auto-ignition temperature 323°C. We also need to keep the temperature as low as possible, as there has been evidence in bench tests that LAB can be heat damaged.

The design flow rate was set at 19.3 LPM LAB (1000 kg/hr) due to the logistics of rail tanker delivery underground. Six rails cars at 2.2 tonnes each can be delivered underground per day, for total of 79.2 tonnes/week for six days. This delivery rate can be purified for AV fill with the plant operating 3.5 days at 24 hrs/day or over six days/week with two operating shifts/day. This processing rate is also limited by the power and cooling requirements, where this flow rate will require about 250kW electrical power and cooling (including stripping). The plant design incorporates a boiling water cooling loop for condenser heat recovery, which is used to pre-heat the feed, resulting in net power requirement of less than 180kW after warm up. To obtain 99% PPO/LAB separation six stages are required. A significant design constraint was to minimize the column dP in order to reduce the reboiler temperature. For the column internals we selected dual-flow trays, rather than the more traditional down-comer and valve tray arrangement. The dual-flow trays offer higher efficiency per stage, since phase contact occurs throughout the stage volume rather than restricted to the liquid zone, as with down-comers. While down-comers would have required less height per stage, more stages would have been required, resulting in increased pressure and temperature at the reboiler. The typical benefits of down-comers, which are tolerance for fouling and tray weeping, and good turn-down capacity, are less important for our application. However, it is critical that the tower is fabricated to precise dimensions and installed level for the dual-flow trays to perform correctly. To eliminate bolted hardware within the column, for reduced particulate, we designed a unique custom tray hold-down system utilizing electro-polished ½" SS316L cotter pins. Using customized tooling, the pins were flared open from below during installation, which provides a very tight tray hold-down to minimize tray vibration and weeping on the trays.

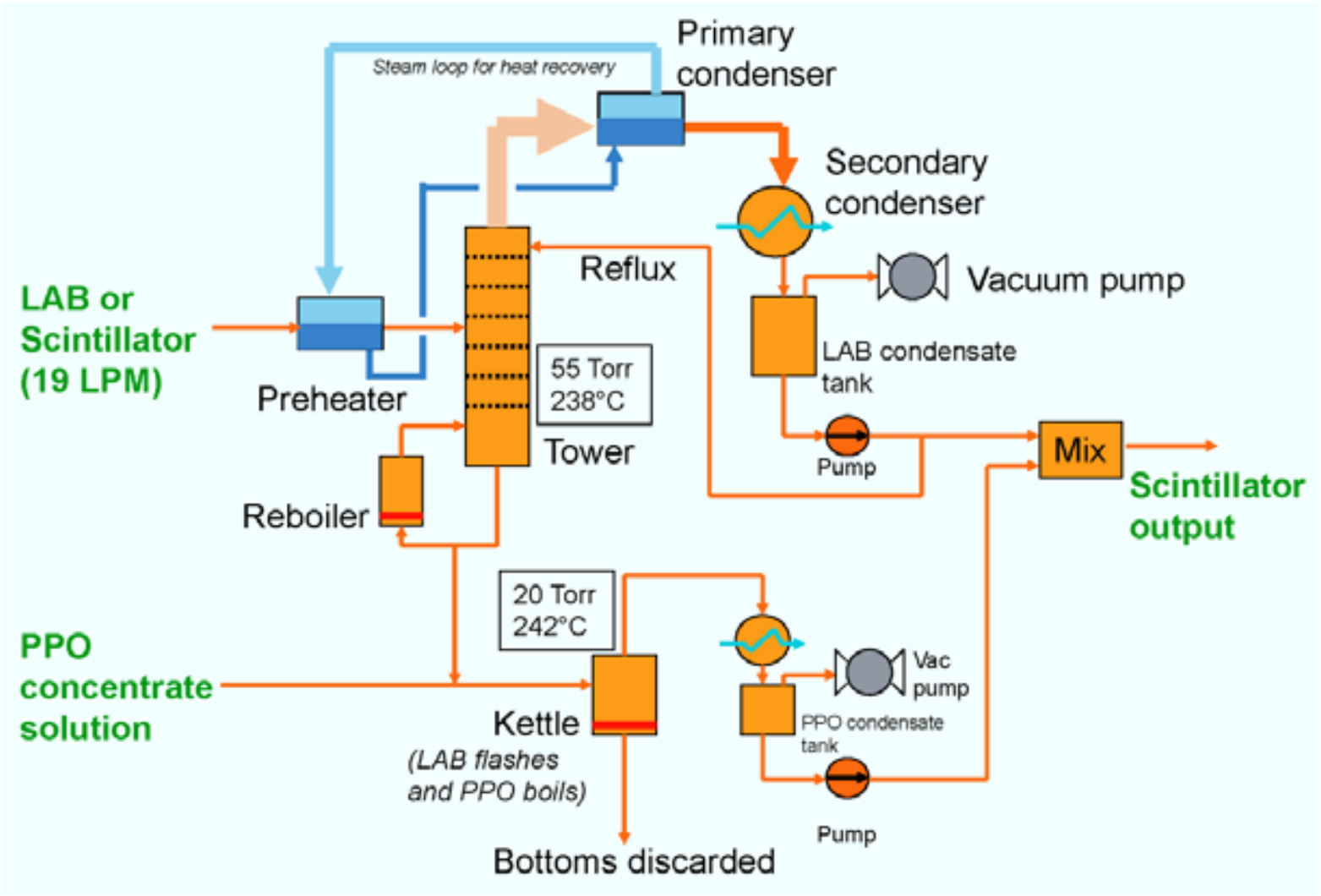

**FIGURE 2.** Simplified process flow diagram for the SNO+ distillation system. The distillation tower is 32" dia x 13'-7" H six-stage with dual-flow trays designed for vacuum distillation at 55 Torr for LAB or scintillator at 19 LPM. Also, in parallel a single-stage kettle flash distills a concentrated stream of PPO dissolved in LAB at 120 g/L. For distillation of scintillator the tower bottoms are processed through the flash kettle for recovery and purification of the PPO (dual-stream distillation).

The PPO will also be purified by distillation, with the process operating at 242°C and 20 Torr vacuum. The PPO is first dissolved into the LAB as a concentrated solution so that it can be fed into the distillation as a liquid. At these conditions the LAB will "flash" distill, and recombine with the PPO in the condenser in order that the PPO remains in solution. If scintillator is fed into the main tower then the bottoms will be concentrated PPO, which can be fed into the PPO distillation kettle for parallel repurification and recombination with the LAB (see fig. 2). We refer to this as the "dual-stream" distillation, and it provides a method for repurification of the scintillator on-line on the AV.

## Solvent-Solvent Extraction with Water

Solvent-solvent extraction is an equilibrium stage process where two immiscible solvents are brought into close phase contact, and then allowed to re-separate. When water is the extract solvent we refer to the process as water extraction, and the efficacy is based on significant difference in solubility between water and LAB. At each theoretical stage a chemical species $i$ has distribution coefficient $K_i = x^r_i/x^e_i$ which is the relative solubility where $x_i$ is the molar concentration between the raffinate (LAB) and the extract (water). The multistage process is implemented as a column where the heavy and light phases flow *counter-current*, with a dis-engagement stage either at the top above the heavy phase feed, or at the bottom below the light phase feed. There are a variety of technologies for the column internals, such as packing or mechanical agitation, to realise efficient phase mixing according to desired process conditions. Laboratory bench testing has shown water extraction to be highly effective [5] at removing ionic heavy metal species such as U, Th, Ra, K, and Pb, and will also remove suspended ultrafine particulate, which tends to be charged. $K_i$ is typically a strong function of temperature, and the process is often best heated. However the process is not effective in removing optical impurities, which tend to be non-polar oxidized organics which have greater affinity for the organic solvent. Thus the water extraction is intended for on-line polishing re-purification after detector fill, to reduce backgrounds from AV leaching and Rn plate-out daughters.

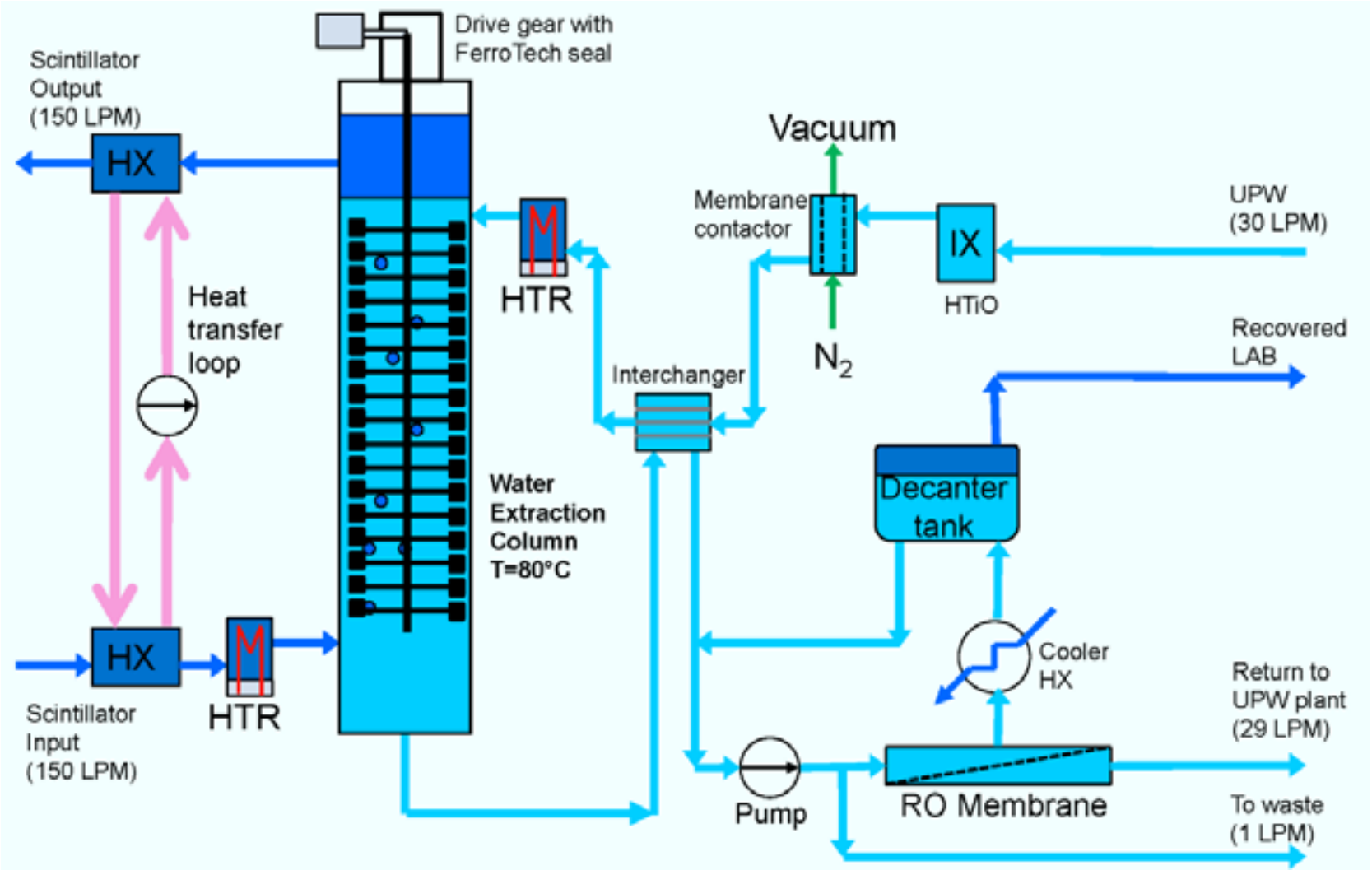

**FIGURE 3.** Simplified process flow diagram for the water-extraction scintillator purification column. The column is a 30" Dia x 18'-4" H 22-stage Scheibel™ column [6], designed to process 150 LPM raffinate (scintillator) and 30 LPM extract (water).

The process is designed to purify at the flow rate of 150 LPM for scintillator. This design basis is to allow the entire detector volume (910 $m^3$) to be processed within 100 hours (~4 days), to allow the detector volume to be purified in quasi-batch mode (minimize mixing of the re-purified scintillator) using temperature regulation. Also for the repurification to be effective, the turnover time should not be longer than the half-lives for $^{224}$Ra and $^{222}$Rn (~4 days) if these are internally supported. A greater challenge was to minimize the ultrapure water requirement, limiting UPW flow to 30 LPM, which is supplied from the SNOLAB UPW plant in parallel with purification of the cavity and PSUP water. The design of the column is to maximize stages, since the $K$ values may not be large, and the partition per stage potentially low due to restricted water flow. A further constraint is the vertical height available for a column within the plant (about 20 feet), which limits number of stages within the column design. To maximize the number of stages a Scheibel™ column [6] is used, which uses a rotating impeller stack with 22 rotating stages and baffle plates. The LAB feed enters the bottom of the column and moves up to the outlet at the top of the column under pressure from the driving pump. The water enters the top of the column and flows down against the prevailing LAB flow due to gravity and higher density. Disengagement occurs in the column sump, which is instrumented with a capacitance probe for interface level measurement. The LAB outflow is controlled by the column pressure, while the water outflow is controlled off the interface level. For increased efficiency the column is designed to operate at 80°C, which then requires heating and cooling power for the process. As for the distillation, the process is only feasible if the process heat can be recovered, due to power and cooling constraints.

A simplified process flow diagram is shown in Fig 3. Heat from the scintillator output is recovered in a transfer circuit to preheat the feed, where each of the HX's are twin 100 kW tube-in-shell exchangers, and the temperature is then boosted with a 130 kW kettle heater. Similarly on the water side the heat is recovered through a direct interchanger (62 kW) with water temperature then boosted with a 165 kW kettle heater. The water extraction process at 150 LPM requires approximately 340 kW power during startup, and 240 kW power and cooling at operating temperature. The UPW feed is polished with hydrous titanium oxide column (HTiO) for the removal of Ra, Th, and Pb, and a degassing membrane for removal of Rn and $O_2$. The spent water is processed through a low-fouling RO membrane (with 5x 8"x40" units) to pre-clean and reject entrained and dissolved LAB before returning the water to the UPW plant. The RO cross-flow loop includes cooling and a decanter tank for accumulation of entrained LAB/scintillator. The water extraction must be followed by stripping to remove residual water.

The water extraction column and decanting loop can be used also to separate the Te (or Nd) from the scintillator to unload and recover the double beta decay material.

## Gas Stripping with Steam and Nitrogen

Gas stripping is also a counter-current staged equilibrium process, in this case based on difference in the vapour-liquid partitioning of a volatile species versus temperature (or equivalently the partial pressures represented by the Henry coefficients). The process is effective for impurities with large vapour pressures, such as gases and volatile liquids. Typically the feed is introduced to the top of a column filled with high surface area packing. A stripping gas, such as nitrogen, is input at the bottom of the column and driven up the column through the packing to exhaust at the top. The gas flow can be driven by pressure, or by a vacuum pump at the exhaust. The liquid phase falls down the column by gravity, with several re-distributor trays to ensure uniform flow through the packing.

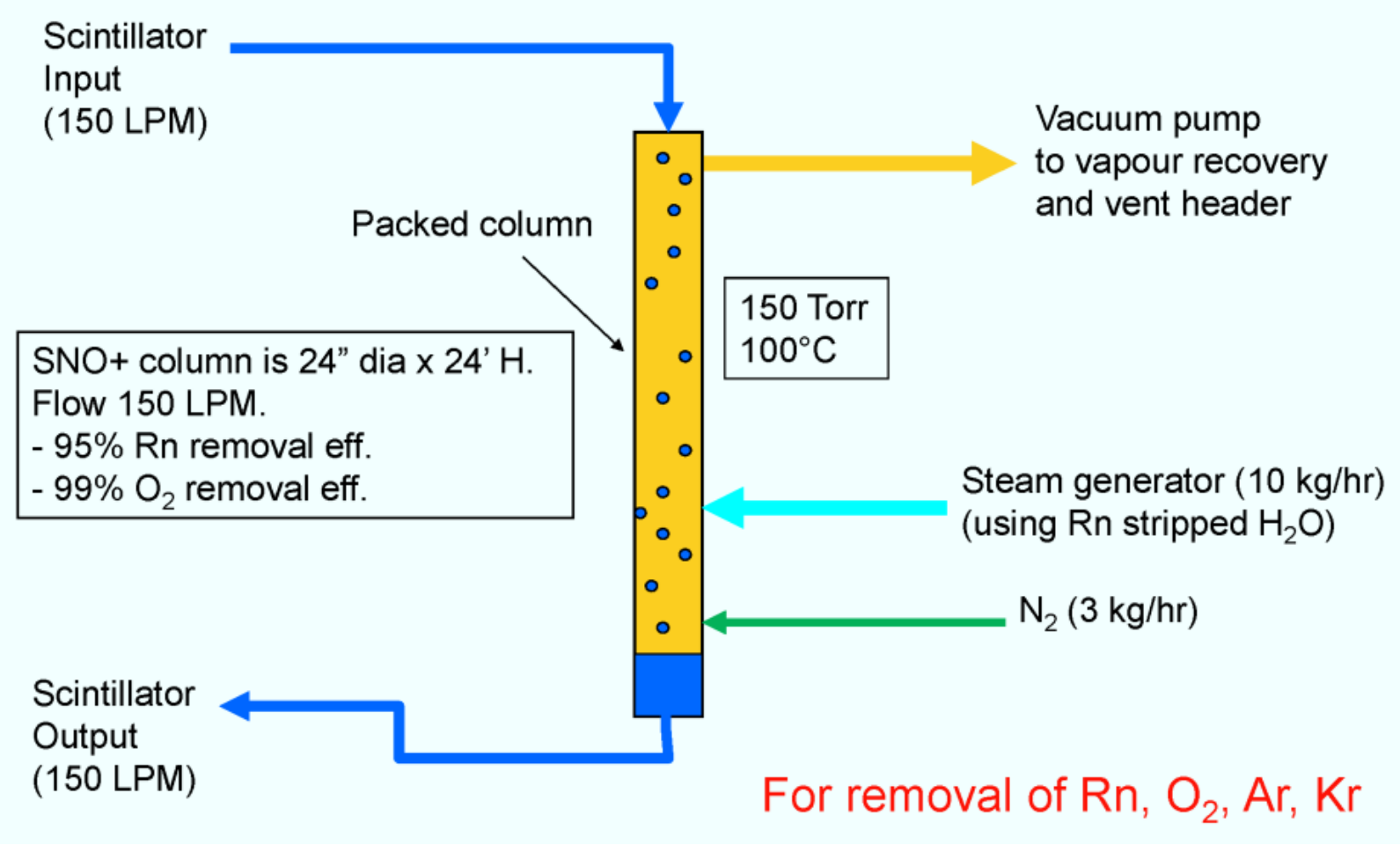

**FIGURE 4.** Simplified process flow diagram for the gas stripping process. The column is a 24" Dia x 22'-8" H with 19 elements of Koch-Glitsch FlexiPac high density SS316L packing design to purify scintillator at 150 LPM and 100°C. Stripping gas is combination of super-heated steam (10 kg/h) and nitrogen (3 kg/h). Heat recovery HX's, kettles, and steam generator not shown.

For SNO+ the volatile impurities of interest are Rn, Ar, Kr for radioactive backgrounds, and oxygen, which will degrade the optical purity and also degrades the beta-alpha discrimination. The stripping is also important to remove residual water from the scintillator, which can cause increased light scattering or even optical opaqueness. However, being non-polar the solubility of noble gasses is very high in aromatic solvent, for example the Henry coefficient for Rn in LAB is about ~11 atm/mf [9], so that the Rn partitions about 87% into the LAB at 1 atm. This could be overcome by more stages in the column, however this is constrained by the overhead height available in the plant. The stage effectiveness can be increased by reducing the pressure to vacuum, however the number of effective stages with the column is reduced at low pressure. A significant design challenge was that it is highly impractical to use only nitrogen as the stripping gas, due to difficulty shipping liquid nitrogen underground in large quantity. Consequently the column is designed to use primarily steam (10 kg/hr) and nitrogen (3 kg/hr). Steam is effective because the Henry coefficient for Rn in water is low, and since the UPW plant effectively degasses Rn from the water [2]. The steam generator includes a super-heater to ensure the steam is dry under vacuum at 100°C.

Based on column simulations by KMPS, the column is designed to operate at 100°C and 150 Torr vacuum, which results in approximately three equivalent stages and 95% efficiency for Rn reduction. The process flow rate is 150 LPM, as the stripping process will follow the water extraction. The column will have higher efficiency for $O_2$, Kr, and Ar as the solubility of these gasses in LAB is lower [10]. The water extraction process at 150 LPM requires approximately 160 kW during startup, and 140 kW at operating temperature.

## Functional Metal Scavengers

Coming primarily from the pharmaceutical industry there is a considerable range of functional metal scavengers now commercially available, which are developed for removing heavy metals from aqueous or aromatic solvents. Various scavenger products from Reaxa LTD were extensively investigated as reported in [5]. QuadraSil-AP™ [11], which has an aminopropyl functional group on a 50μm mesh silica gel, was found to be highly effective for Pb and Ra removal. The motivation for developing the scavenger purification is to provide an alternate and complementary process to water extraction, that is also expected to be effective for Bi. A feature of QuadraSil-AP is that it can be regenerated with HCl acid, and the metals recovered and analyzed with coincidence beta-alpha counting. This provides a method for *ex-situ* radioactivity assay for $^{224}$Ra and $^{226}$Ra, as was done in SNO [12]. The method may also allow separate determination for $^{210}$Pb and $^{210}$Bi, to overcome the spectral degeneracy of $^{210}$Bi with CNO neutrinos.

The challenge was the design of a process column, as in industry the scavengers are mixed batch-wise and recovered by filtration. In small scale testing, tall columns are more efficient, with 95% removal efficient for Pb at flow rate up to 50 bed-vol/hr for a column with height/dia = 45 [5]. However, the column dP scales with height, and can become very large leading to restricted flow and channeling effects. The process as designed has six columns in two banks of three, with columns 6" diameter and 180" high for the bed volume (83 L). If filled to 100%, the 150 LPM flow corresponds to 36 bed-vol/hr for three columns, where 2x three columns can be operated in series. If half filled, at 70 bed-vol/hr, the efficiency for $^{212}$Pb and $^{224}$Ra spike removal is still over 90% [5].

## FLUID HANDLING SYSTEM

The fluid handling system provides the external connections to the purification plant for the detector process operations. See figure 5 for simplified schematic of the entire fluid handling system for LAB and scintillator. The system includes the surface transfer facility (STF), the underground transfer facility (UTF), the storage facility (60-tonne tanks), the deck valve cabinet (AV flow control), Te-LAB loading process, and the PPO-LAB preparation process. The STF is the surface delivery terminal which has a two-way pumping station and a 70-tonne tank, to enable flexible logistics for receiving 22-tonne road tanker trucks with LAB, and subsequently filling the six 2.2-tonne mine rail tanker trucks for shipment underground. The UTF is the underground equivalent of the STF, and is located in the SNOLAB outer carwash room for the unloading or loading of the rail tanker trucks and transfer of the LAB to the 60-tonne tanks. The 60-tonne tanks provide buffer capacity for unloading the LAB rail cars while the distillation for filling the detector runs continuously, with capacity for about 3.5 days of feed. The tanks would also provide the option to store the PPO master solution and/or the Te-LS concentrated solution if required.

The PPO-LAB preparation equipment is for making and pre-purifying a concentrated solution of PPO in LAB at 120 g/L. The PPO is mixed as a concentrated solution as it requires purification before blending into the detector. The equipment consists of an 840L mixing tank under cover gas, a hopper vessel with $N_2$ purge, and connections for purified LAB and ultrapure water. It is known that the PPO as delivered contains significant potassium and insolubles removable by batch water-extraction within the tank. Each PPO-LAB batch is then filtered at 5μm and 0.2μm, before sent to the flash distillation described above. The Te-LS preparation equipment consists of an 880L tank under cover gas, with variable drive mixer, recirculation, and ingredient tanks with feed controllers for LAB, telluric acid, and the loading agent(s).

The purification plant is designed with a main header line from the recirculation pumps, where the purification processes branch off and then return to the header. The configuration then allow one purification mode (ie. scavenger, distillation, or water-extraction), followed typically by stripping. The recirculation to the AV includes a final stage cooler (to 12°C), filtration (0.05μm), and flow control to maintain level in the AV. Level control for the AV is quite critical and must maintain zero dP at the bottom of the AV neck to with 6". The flow to the AV is decoupled from the process pressure through a high point head tank at atm pressure to ensure stable flow control.

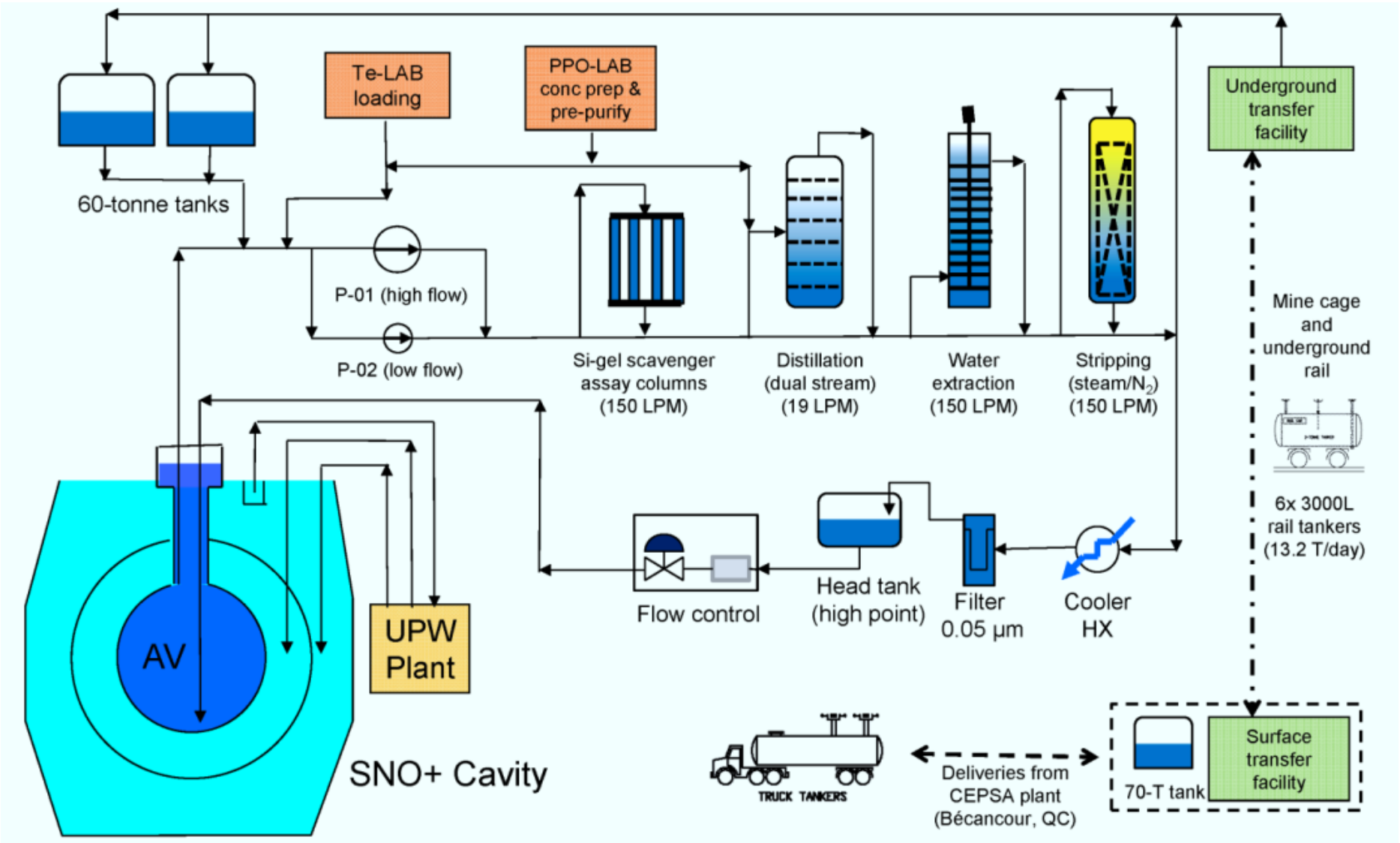

**FIGURE 5.** Simplified process flow diagram for the SNO+ scintillator liquid handling. Major plant areas are the surface transfer facility, underground transfer facility, 60T tanks storage area, scintillator purification plant, and the AV deck valve cabinet.

## CONCLUSIONS AND CURRENT STATUS

A scintillator purification and liquid handling system has been designed and constructed for the SNO+ experiment. It features distillation of the LAB and PPO at up to 19 LPM fill rate, and includes $N_2$/steam stripping for removal of Rn and $O_2$. For on-line repurification there is water extraction or scavenger silica gel columns, followed by gas stripping, which can repurify the scintillator at 150 LPM for complete detector turnover within 100 hours.

The plant is designed and built from SS316L with electro-polished internals and vacuum tight seals. The plant has been 100% helium leak tested. Passivation and high-purity cleaning of the plant is currently in progress, and soon we will begin the plant commissioning stage, and a HAZOP safety review of plant operability.

## ACKNOWLEDGMENTS

Capital construction funds for SNO+ is provided by the Canada Foundation for Innovation (CFI) and this work was supported by the Natural Sciences and Engineering Research Council of Canada (NSERC). We are grateful to SNOLAB for hosting the SNO+ experiment and for on-going support in the construction and R&D. We thank Vale and the staff at Creighton mine for their helpful support of SNOLAB and the SNO+ experiment.